\documentclass[conference]{IEEEtran}
\IEEEoverridecommandlockouts
\usepackage{cite}
\usepackage{amsmath,amssymb,amsfonts}
\usepackage[ruled]{algorithm2e}
\usepackage{algorithmic}
\usepackage{graphicx}
\usepackage{textcomp}
\usepackage{xcolor}
\usepackage{url}
\usepackage[colorlinks, linkcolor=purple, citecolor=purple]{hyperref}
\usepackage{multirow}
\usepackage{makecell}
\usepackage{booktabs}

\def\BibTeX{{\rm B\kern-.05em{\sc i\kern-.025em b}\kern-.08em
    T\kern-.1667em\lower.7ex\hbox{E}\kern-.125emX}}
\begin{document}

\title{FPHammer: A Device Identification Framework based on DRAM Fingerprinting
}

\author{
 \IEEEauthorblockN{Dawei Li, Di Liu, Yangkun Ren, Ziyi Wang, Yu Sun, Zhenyu Guan\IEEEauthorrefmark{1}, Qianhong Wu, and Jianwei Liu}
 \thanks{\IEEEauthorrefmark{1} Corresponding author. This paper has been accepted by TrustCom-2023.}
 \IEEEauthorblockA{School of Cyber Science and Technology, Beihang University}
 \IEEEauthorblockA{\{lidawei, liudi2020, ren1319, wangzy112, sunyv, guanzhenyu, qianhong.wu, liujianwei\}@buaa.edu.cn}
}


\maketitle

\begin{abstract}
The device fingerprinting technique extracts fingerprints based on the hardware characteristics of the device to identify the device.
The primary goal of device fingerprinting is to accurately and uniquely identify a device, which requires the generated device fingerprints to have good stability to achieve long-term tracking of the target device. 
However, the fingerprints generated by some existing fingerprinting technologies are not stable enough or change frequently, making it impossible to track the target device for a long time.
In this paper, we present FPHammer, a novel DRAM-based fingerprinting technique. 
The device fingerprint generated by our technique has high stability and can be used to track the device for a long time. 
We leverage the Rowhammer technique to repeatedly and quickly access a row in DRAM to get bit flips in its adjacent row. 
We then construct a physical fingerprint of the device based on the locations of the collected bit flips.
The evaluation results of the uniqueness and reliability of the physical fingerprint show that it can be used to distinguish devices with the same hardware and software configuration.
The experimental results on device identification demonstrate that the physical fingerprints engendered by our innovative technique are inherently linked to the entirety of the device rather than just the DRAM module.
Even if the device modifies software-level parameters such as MAC address and IP address or even reinstalls the operating system, we can accurately identify the target device.
This demonstrates that FPHammer can generate stable fingerprints that are not affected by software layer parameters.

\end{abstract}


\section{Introduction}
The device fingerprinting technique realizes device identification by collecting specific attributes of the target device. This technique can be used constructively as well as destructively. 
Examples of constructive use include device authentication \cite{alaca2016device}, web bot detection \cite{jonker2019fingerprint}, etc. 
Examples of destructive use include attackers tracking targeted devices based on fingerprints \cite{englehardt2016online}.
Regardless of the usage scenario, the key challenge is to generate a fingerprint that can uniquely identify the device, 
and the generated fingerprint needs to be highly stable so that the target device can be identified and tracked.

Browser fingerprinting is a stateless device identification technique. 
This technique usually equates identifying a browser with identifying a user and builds a fingerprint by collecting the attributes of the target device's browser and system \cite{laperdrix2016beauty}. 
Device fingerprints are generated every time a user visits a website. 
For the same user, user tracking is achieved by linking the fingerprints generated multiple times to the same user.
Relying solely on browser properties and system configuration information to uniquely identify a device is not accurate enough. 
Some works \cite{laor2022drawnapart, sanchez2018clock} combine the hardware characteristics of the device to form device fingerprints, which can increase the accuracy of device identification.

A significant challenge of browser fingerprinting is that the stability of the generated device fingerprint is not high so the device cannot be identified by a simple fingerprint matching method, but a complex method such as machine learning needs to be used to identify the device.
This is because most of the attributes of the browser will change over time, which may be caused by some user operations or system upgrades.
This results in the inability to uniquely identify devices with the same hardware and software configuration. 
This also leads to shorter tracking times for devices, with a median tracking time of less than two months in this case, according to Vastel et~al. \cite{vastel2018fp}
Additionally, machine learning-based device identification methods take longer and lead to a dilemma. 
That is, the training of the model needs to collect enough fingerprints of the same device, but linking the fingerprints to the same device requires a trained machine learning model.

To address this challenge, we propose FPHammer, a physical fingerprint generation technique based on Dynamic Random-Access Memory (DRAM) in this paper. 
Random variations introduced in the manufacturing process of the same hardware components can be used to extract a device's unique fingerprint. 
The physical fingerprints generated by our technique have high stability. 
Therefore, fingerprints belonging to the same device can be linked through a simple fingerprint matching algorithm without resorting to more complex machine learning algorithms.

Our proposed fingerprint generation technique can be used in scenarios where attackers actively detect target devices. 
In this way, the attacker remotely executes the source code on the target device and returns the generated fingerprint, or combines the fingerprint with browser properties for browser fingerprinting scenarios.
For constructive use, administrators can authenticate devices based on the device's unique physical fingerprint.
In addition, it can be applied to device-based paid service scenarios, which require software to only run on specific devices.

\textbf{Our Contribution.} 
We design and implement FPHammer for extracting physical fingerprints from a device's DRAM components, based on the Rowhammer technique.
By repeatedly and quickly accessing a row in DRAM to get bit flips in its adjacent rows, we form a physical fingerprint based on the location of the bit flips.
We evaluate the performance of our DRAM fingerprinting technique through a series of experiments, and the results show that the generated physical fingerprints have high stability.
It can thus be used to distinguish devices with the same configuration and enable long-term tracking of the target device.

In summary, the main contributions of this paper include:
\begin{itemize}
    \item We propose a device identification framework based on DRAM fingerprinting, and propose a device identification algorithm for linking fingerprints belonging to the same device.
    \item We design and implement a Rowhammer-based fingerprinting technique that can generate device fingerprints at runtime on a personal computer running the Linux operating system and equipped with DDR4 memory.
    \item We evaluate the properties of the generated fingerprints, and the results show that the generated fingerprints can be used to distinguish devices with the same hardware and software configuration.    
    \item We conduct several experiments in the laboratory scenarios. 
    The experimental results show that even if the target device changes virtual attributes such as MAC address and IP address, or even reinstalls the operating system, we can still accurately identify and track the target device through physical fingerprints.
\end{itemize}


\section{Background}

Rowhammer is a DRAM vulnerability in which when a row (\textit{aggressor row}) in the memory is repeatedly and rapidly accessed, some bits of its adjacent rows (\textit{victim~rows}) will be flipped \cite{kim2014flipping, van2016drammer, gruss2018another,cojocar2020we,de2021smash,jattke2022blacksmith}.
The root cause of this phenomenon is that repeated access to a certain DRAM row will accelerate the charge leakage of adjacent rows \cite{kim2014flipping}.
To achieve rapid access to the rows in the DRAM, the cache mechanism between the CPU and the memory needs to be bypassed.
Commonly used methods include flushing the cache using \textit{clflush} or \textit{clflushopt} instructions \cite{cojocar2020we}, cache eviction \cite{aweke2016anvil, gruss2016rowhammer}, uncached memory \cite{qiao2016new}, and so on.

To successfully induce bit flips in DRAM, the first step is to get the mapping relationship between virtual addresses to physical addresses and then to DRAM addresses (a specific channel, DIMM, rank, bank, row, and column).
Then, we need to determine the appropriate hammering patterns to implement reliable bit flips in DRAM.
There are mainly three ways to trigger bit flipping. The first way is the native code execution, which is running source code on a personal computer or mobile device.
The second way is to trigger by running JavaScript code on the browser \cite{de2021smash, gruss2016rowhammer}.
The third way is to remotely trigger bit flips by establishing a fast network connection between the attacker and the victim and sending carefully crafted packets of network traffic to the victim device, 
such as Throwhammer \cite{tatar2018throwhammer} and Nethammer \cite{lipp2020nethammer}.

\begin{figure}[!t]
	\centering
	\includegraphics[width=3.4in]{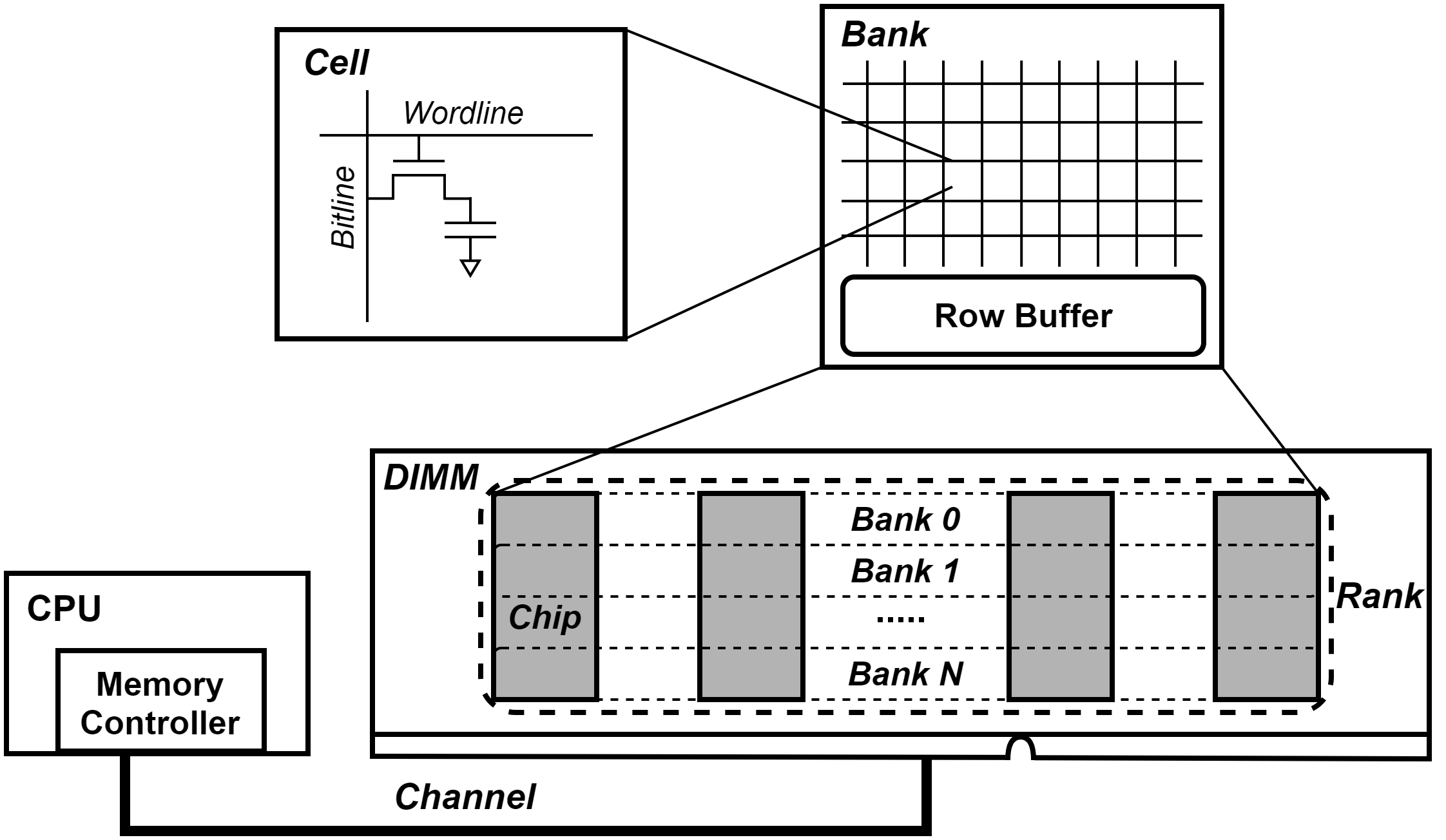}
	\caption{DRAM organization.}
	\label{DRAM-organization}
\end{figure}

\textbf{DRAM organization.} 
As shown in Fig. \ref{DRAM-organization}, the memory controller in the CPU sends a memory request to the DRAM \textit{rank} through the \textit{channel} and one channel can carry multiple DRAM modules (DIMMs).
A DRAM module has one or more DRAM ranks and a rank generally has 8 DRAM \textit{chips}. A DRAM rank has multiple \textit{banks} and one bank spans multiple chips.
DRAM banks are further divided into \textit{rows} and \textit{columns} of memory \textit{cells}, where each cell consists of a transistor and capacitor.
For \textit{true cells}, a fully charged capacitor represents a bit of data “1”, a fully discharged capacitor represents a bit of data “0”, and the opposite is true for \textit{anti cells} \cite{kwong2020rambleed}.
The memory controller is responsible for converting physical addresses into corresponding channels, DIMMS, ranks, and banks in the memory.

\textbf{Reverse Engineer the Address Mapping.} 
The mapping of virtual addresses to physical addresses is implemented through the operating system, such as accessing the \textit{pagemap} interface or using \textit{huge pages}.
However, not all the mapping relationships between physical addresses and DRAM addresses are public.
For Intel's CPUs and AMD's latest CPUs, we need to reverse engineer the memory address mapping function.

DRAMA \cite{pessl2016drama} and DRAMDig \cite{wang2020dramdig} are two software-only DRAM address mapping reverse engineer tools, and we use DRAMA to get the address mapping function in our evaluation experiments. 
They all use the \textit{time side channel} to determine the bank bits. 
Specifically, each bank in DRAM has a row buffer, which is used to access a row in the bank.
Randomly select some address pairs, access them repeatedly and alternately, and measure the average access time. 
If a pair of addresses are in the same bank, their access time will be significantly higher than those address pairs that are not in the same bank.
This phenomenon is also called \textit{row buffer conflicts}, so we can find an address set whose addresses are all in the same bank and determine the bank bits in the address.

\textbf{Hammering patterns.} 
Hammering patterns refer to the combination of different aggressor rows, which have a great influence on the probability of bit flips. By selecting a suitable hammering pattern, more bit flips can be obtained. 
There are currently five main hammering patterns:
\begin{itemize}
    \item \textbf{One-location Rowhammer and Single-sided Rowhammer.}
    One-location hammering pattern, which repeatedly accesses a single row in the bank. This pattern only re-opens one row repeatedly, so it does not cause row buffer conflicts \cite{gruss2018another}.
    Although this pattern induces a few bit flips, it can bypass the protection mechanism based on memory access pattern analysis in some cases.
    Single-sided hammering pattern just repeatedly accesses two rows in the same bank, and there is no specific positional relationship between the two rows.

    \item \textbf{Double-sided Rowhammer.}
    Double-sided hammering accesses two rows that satisfy the adjacency relationship. 
    The two rows are adjacent to the same victim row, which can increase the probability of inducing bit flips.
    For most DDR3 DRAMs that lack in-DRAM \textit{Target Row Refresh (TRR)} mitigation mechanisms, single-sided or double-sided hammering can induce bit flips. 
    But for the new DDR4 DRAM with TRR mechanism, it is difficult for these two patterns to induce bit flips in practice.

    \item \textbf{Many-sided Rowhammer.}
    This pattern has the same adjacency relationship between two aggressor rows as the double-sided pattern.
    Frigo et~al. \cite{frigo2020trrespass} first presented a many-sided Rowhammer tool, which can automatically identify hammering patterns and induce reliable bit flips on DDR4 DRAM with in-DRAM TRR mitigation.
    Since the TRR sampler can only track a limited number of aggressor rows, once the number of aggressor rows exceeds the size of the TRR sampler, the attacker may successfully induce bit flips.

    \item \textbf{Non-uniform Rowhammer.}
    In the above four patterns, the access frequency and other parameters of aggressor rows are uniform, which is easier to be evaluated by the TRR mechanism. 
    The non-uniform pattern proposed by Jattke et~al. \cite{jattke2022blacksmith} makes it difficult for TRR to accurately estimate potential aggressor rows by randomizing the three temporal parameters of the aggressor row.
    According to their experiments on personal computers, this pattern got bit flips on all 40 DDR4 DIMMs from the three major DRAM suppliers with a 100\% success rate.
\end{itemize}

Moreover, some previous works \cite{kim2014flipping, van2016drammer} have shown that most of the bit flip locations in DRAM are stable, and random differences in the manufacturing process make the bit flip locations different in different DIMMS of the same design.
Therefore, the flipped bits produced by the Rowhammer technique can be used to construct a unique physical fingerprint of the device.

\section{DRAM-based Device Identification Framework}

\subsection{Framework Overview}
\begin{figure}[!t]
	\centering
	\includegraphics[width=3.4in]{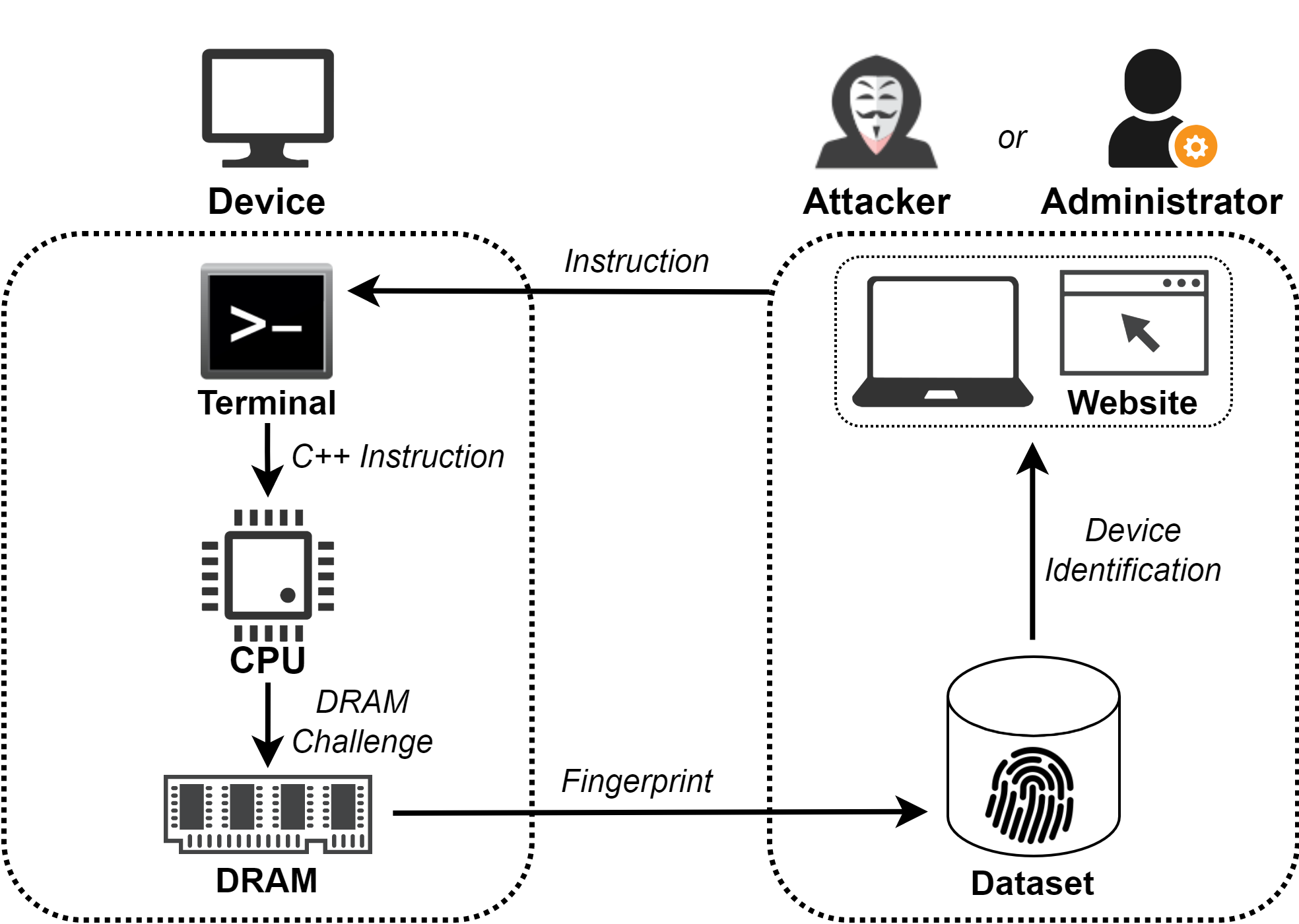}
	\caption{The overview of device identification framework.}
	\label{identification-framework}
\end{figure}

As illustrated in the Fig. \ref{identification-framework}, our proposed framework consists of two components.
The first part is the generation of the target device fingerprint, the goal is to generate a stable physical fingerprint on the target device that can uniquely identify the device.
We generate the unique physical fingerprint of the device based on the memory module in the device. The generated fingerprint is directly bound to the hardware of the device and is not affected by the system and software. 
More specifically, we use the Rowhammer technique to get bit flips in DRAM and form the fingerprint of the device based on the location of the bit flips.

The second part is to identify and track the device. 
The attacker can collect fingerprints by actively detecting the target device, or when the target device accesses the website built by the attacker.
Administrators can ask the target device to run code that generates a physical fingerprint and uses the device's fingerprint to identify the device or authenticate the device's legal identity.
The attacker or administrator links fingerprints belonging to the same device to the same device identifier and stores all fingerprints in the dataset.
For the newly collected fingerprints, we match the new fingerprints with the fingerprints in the dataset according to the identification algorithm. 
If the match is successful, it means this is a known device in the dataset, and we link the fingerprint to the corresponding device identifier. 
Otherwise, this is a new device, and we assign a new device identifier to the fingerprint.

\subsection{Threat Model}
Our proposed framework involves two parties: the target device and the entity that identifies the target device.
We consider two realistic scenarios of device identification, where the entity identifying the target device is an attacker or an administrator, respectively.
Entities in different scenarios have different capabilities.
We assume that the target device is susceptible to Rowhammer bit flips and is not equipped with ECC memory.

\textbf{Scenarios 1: Destructive Use For Attacker.} 
The attacker's goal is to obtain the unique physical fingerprint of the target device and keep track of the device.
For an active attacker capable of actively detecting the target device, we assume that there is a sufficiently fast network connection between the attacker and the victim device, 
and the attacker can remotely obtain the physical fingerprint generated by the victim device. 
Attackers can leverage the Throwhammer \cite{tatar2018throwhammer} and Nethammer \cite{lipp2020nethammer} tools to remotely trigger bit flips by sending crafted packets of network traffic containing fingerprinting code to the victim device.
In this way, a Single-sided Rowhammer or Double-sided Rowhammer can be realized. 

\textbf{Scenarios 2: Constructive Use For Administrator.} 
The administrator's goal is to verify that the identity of the target device is legitimate. 
In this case, the administrator requires the target device to run the administrator-provided fingerprint generation code with user privileges and return the fingerprint to the administrator for identification.
If the fingerprints match successfully, the legal identity of the target device is verified.
This unique physical fingerprint based on device hardware can be used as an authentication root, which is suitable for scenarios where the identity of physical devices needs to be strictly authenticated.
For example, business scenarios where billing and authorization of services are based on specific devices, and scenarios where software is only allowed to run on specific devices.

\subsection{Device Identification Algorithm}
When a new fingerprint $f_u$ is collected, we need to determine whether $f_u$ matches the fingerprint of an existing device in the dataset, or comes from a new device.
For the former case we link $f_u$ to an existing device $id$, for the latter case we create a new device identifier for $f_u$.

For a certain device, we identify it as $id_i$. The $k$ fingerprints belonging to this device are represented by $f_{ij},(j=1,...,k)$.
Given a known dataset of device fingerprints $F$ and a newly collected fingerprint $f_u \notin F$. 
If $f_u$ matches the fingerprint set $<f_{ij}>$ of a certain device $id_i$, the device identification algorithm returns the corresponding device $id_i$. 
If no device fingerprint matching $f_u$ is found, it assigns $f_u$ a new $id$.

The device identification algorithm we designed is as described in Algorithm \ref{device-identification-alg}.
We first explain several functions involved in the algorithm, and then describe the specific algorithm flow.
The \textit{FingerprintMatch} function takes two fingerprints as input, evaluates the similarity of the two fingerprints, and returns true if the similarity is greater than a certain threshold and false otherwise.
The \textit{GenerateNewID} function is used to assign a new device $id$ to the fingerprint $f_u$ that does not match the device fingerprints in the existing dataset $F$.
The \textit{GetSimilarity} function takes the fingerprint $f_u$ and the fingerprint set $<f_{ij}>$ corresponding to the known device $id_i$ as input, calculates the similarity between $f_u$ and $<f_{ij}>$, and returns the similarity value.
The \textit{GetRank} function takes the $similarity$ list as input, sorts the elements in the list from large to small, and returns the sorted $similarity$ list.

The detailed device identification process is described as follows.
First, select a fingerprint such as $f_{i1}$ from the fingerprint set $<f_{ij}>$ corresponding to each device $id_i$ to perform fingerprint matching with $f_u$. 
The reason why only one fingerprint is selected for matching here is to improve the matching efficiency.
If $f_{i1}$ matches $f_u$, add it to the candidate list $candidates$.
If the list $candidates$ is empty, meaning that $f_u$ does not match the fingerprints in the existing dataset $F$, then we assign $f_u$ a new device $id$.
If there is only one candidate in the list $candidates$, then we link $f_u$ with the device $id$ corresponding to the candidate, and add $f_u$ to the fingerprint set corresponding to this device.
If there is more than one candidate in the list $candidates$, then we compute the similarity between $f_u$ and the set of device fingerprints $<f_{ij}>$ corresponding to each candidate using metrics such as $Jaccard'~index$.
Then we sort the elements in the list $candidates$, link $f_u$ to the device $id$ corresponding to the fingerprint with the largest similarity value of $f_u$, and add $f_u$ to the set of fingerprints corresponding to this device.

\begin{algorithm}[!t]
    \caption{Device identification algorithm based on fingerprint matching}
    \label{device-identification-alg}
    \SetKwProg{Fn}{Function}{}{}
    \SetKwFunction{FMain}{DeviceIdentification}
    \Fn{DeviceIdentification({$F$,~$f_u$})}{
        $candidates \leftarrow \emptyset$ \;
        \For(){$f_{i1} \in F$}{
            \If(){FingerprintMatch($f_u,~f_{i1}$)}{
                $candidates \leftarrow candidates~\cup <f_{i1}>$ \;
            }
        }
        \uIf(){$|candidates| = 0$}{
            \textbf{return} \textit{GenerateNewID()} \;
        }
        \uElseIf{$|candidates| = 1$}{
            \textbf{return} $candidates[0].id$ \;
        }
        \Else{
            \tcc{$|candidates| > 1$}
            \For(){$f_{i1} \in candidates$}{
                $similarity \leftarrow $ \textit{GetSimilarity($f_u,<f_{i1},f_{i2},...,f_{ik}>$)} \;
            }
            \textit{GetRank($similarity$)} \;
            \textbf{return} \textit{$similarity[0].id$} \;
        }
    }
\end{algorithm}


\subsection{Design Flow}
We first define a set of DRAM initialization parameters, and fix this set of parameters to generate the corresponding DRAM fingerprints. 
Then we give the method to construct DRAM fingerprint and the algorithm to generate DRAM fingerprint.

\begin{figure}[!t]
	\centering
	\includegraphics[width=3.0in]{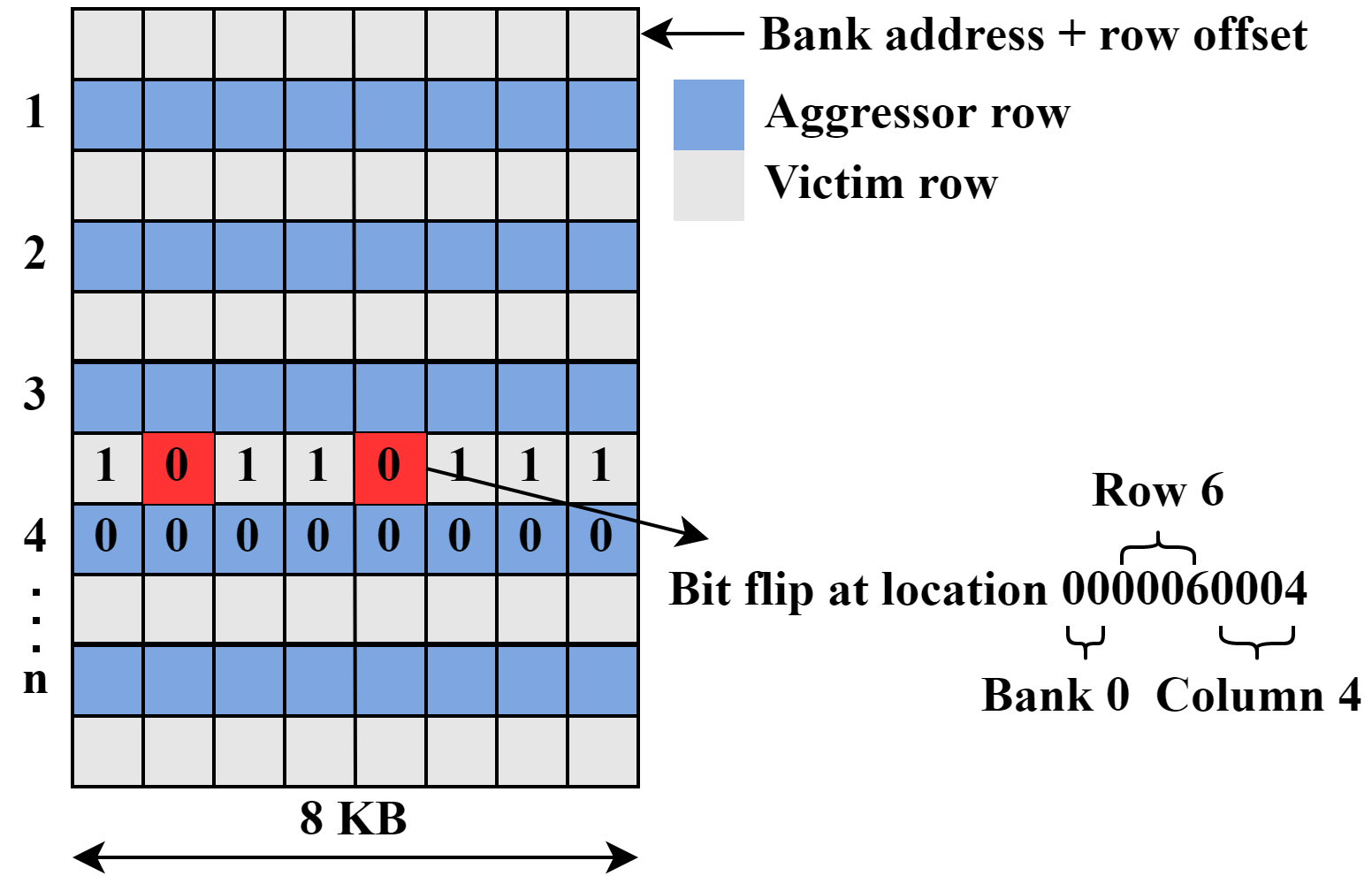}
	\caption{An intuitive example of the selection of DRAM challenges.
    In this example, we chose the many-sided pattern, initialized the aggressor row to 0x00, the victim row to 0xFF, 
    and got bit flips at locations 0000060001 and 0000060004.}
	\label{hammering-pattern}
\end{figure}

\textbf{DRAM Challenge.} 
The DRAM challenge consists of several parameters, each of which affects the number and location of bit flips in the DRAM, and thus determines the resulting DRAM fingerprint.
A more stable fingerprint can be obtained by selecting appropriate challenge parameters.
Fig. \ref{hammering-pattern} shows an intuitive example of the selection of DRAM challenges and the location of bit flips.
We emphasize that the choice of DRAM challenge parameters can be diverse, as long as stable and repeatable fingerprints can be obtained.

\begin{itemize}
    \item \textbf{Row address.}
    The Row address consists of a \textit{bank address} and an row \textit{offset}. 
    The aggressor row and the victim row to be accessed should be located in the same bank. 
    After the specific bank address is determined, the address of the aggressor row is determined by the row offset.

    \item \textbf{Hammering pattern.}
    To get the DRAM fingerprint, valid access patterns need to be used to obtain bit flips under different threat models. 
    For example, in the first scenario of the threat model, an attacker cannot execute code with user privileges, so the \textit{single-sided} pattern and \textit{double-sided} pattern can be applied in this case. 
    In the second scenario of the threat model, the administrator can require the target device to run code with user privileges, so both \textit{many-sided} pattern and \textit{non-uniform} pattern can be applied in this case.

    \item \textbf{Data pattern.}
    The data pattern is determined by the initial values written into the aggressor row and victim row.
    Due to the existence of \textit{true cell} and \textit{anti cell}, this parameter also has a great impact on DRAM fingerprint generation.
    More bit flips can be obtained by choosing appropriate initial values for the aggressor row and victim row.
    For instance, initialize the aggressor row to 0x00 and the victim row to 0xFF, or initialize them to 0x55 and 0xAA respectively.

    \item \textbf{Measuring times.}
    Measuring times includes the number of banks accessed and the number of times each aggressor row is accessed. 
    The row address, hammering pattern, and data pattern should be the same for each bank.
\end{itemize}

\textbf{DRAM Fingerprint.} 
A set of DRAM challenge parameters determines the state of the DRAM fingerprint, and we construct the DRAM fingerprint based on the bit flip locations in the victim row.
More intuitively, an element in the DRAM fingerprint set (such as 00000600004 in Fig. \ref{hammering-pattern}) represents a location where a bit flip occurs.
The format of the location is (\textit{bank, row, column}), indicating which row and column of which bank the bit flip occurred.
The process of generating a DRAM fingerprint is described in Algorithm \ref{fingerprint-generation-alg}.


\begin{algorithm}[!t]
\caption{The process of fingerprint generation}
\label{fingerprint-generation-alg}
    \KwIn{Row address, Hammering pattern, Data pattern, Measuring times}
    \KwOut{DRAM fingerprint}
    Allocate the required memory\;
    \While{m $<$ number of measurements}{
        \While{b $<$ number of banks}{
            Initialize the aggressor rows and victim rows\;
        }
        \While{b $<$ number of banks}{
            Hammer the aggressor rows\;
            Scan the victim rows and output bit flip locations\;
        }
    }
\end{algorithm}

\section{Experiments in the laboratory setting}

\subsection{Property Evaluation}
We tested the reliability and uniqueness properties of the generated DRAM fingerprints on two DIMMs with the same design, the same specifications, and the same production batch.
The CPU model of our test computer is Intel(R) Core(TM) i7-10700 CPU @ 2.90Hz, the CPU architecture is \textit{Comet lake}, and the operating system is Ubuntu 20.04.
The memory is Samsung DDR4 SDRAM without ECC (Error Checking and Correcting) function, which size is 8 GB and frequency is 2932 MHz.
Before running the fingerprint generation code, we first use DRAMA \cite{pessl2016drama} to get the DRAM address mapping function.



\textbf{Jaccard$'$ index.} 
Based on our observations in the evaluation, for the same DIMM, the location and number of bit flips included in different fingerprint queries are not exactly the same for a given DRAM challenge.
When the size of the two sets is very different, the value calculated by the conventional Jaccard index \cite{xiong2016run} cannot accurately reflect the actual situation.
The original Jaccard index is not suitable for identifying target devices.
Therefore, we modify the Jaccard index as follows:

\begin{equation}
    \label{j_intra}
    Jaccard^{'}(S_n,S_d)=\frac{|S_n \cap S_d|}{|S_n|}
\end{equation}

where $S_d$ represents a large set of DRAM fingerprints obtained from the first few queries, such as $S_d=S_1 \cup S_2 \cup S_3$.
$S_n$ represents the fingerprint set of a new query. 

\textbf{Reliability.} 
Reliability measures the similarity between fingerprints generated by the same DIMM for the same challenge in different measurements.
We use index $J_{intra}(S_1,S_2)$ to represent the reliability of DRAM fingerprint.
Due to the measurement noise, each measurement result will not be exactly the same. 
Ideally, the bit flip locations contained in the two sets should be the same, so the ideal value of $J_{intra}$ is 1.
In other words, the higher the similarity between two fingerprints, the more likely they belong to the same device.

\textbf{Uniqueness.} 
Uniqueness measures the difference between fingerprints generated by different DIMMs for the same challenge in different measurements.
We use index $J_{inter}(S_1,S_2)$ to represent the uniqueness of DRAM fingerprint.
Ideally, the locations of bit flips contained in the two sets should not overlap, so the ideal value of $J_{inter}$ is 0.
The lower the uniqueness value between two fingerprints, the more likely they belong to different devices.

\textbf{DRAM Challenge.} 
For each process of fingerprint query, we use the same DRAM challenge settings with parameters shown in Table \ref{DRAM-parameters}.
We chose the many-sided pattern for this evaluation and fixed the first aggressor row offset to row 1 of each bank and measured 5 banks.
Because some DDR4 DRAMs have the TRR mitigation mechanism, some rows with bit flipping may be refreshed in the next measurement without bit flipping.
Therefore, we set the number of aggressor rows to 22 and measured 10 times in each process. 
Since some cells may be \textit{true cells} and some may be \textit{anti cells}, we initialize the victim row and aggressor row to 0x55 and 0xAA, respectively. 
In this way, the states of the cells corresponding to the adjacent rows are reversed, and there are bit flipping conditions for the two possible cell situations.

\begin{table}[!t]
    \caption{Parameters that constitute the DRAM challenge.}
    \label{DRAM-parameters}
    \centering
    \begin{tabular}{c c}
    \toprule
    Parameter & Value \\
    \hline
    Row address       & \makecell{bank address = 0 - 4, \\ the first aggressor row offset = 1} \\
    Hammering pattern & 22 sided pattern   \\
    Data pattern      & victim row = 0x55, aggressor row = 0xAA \\
    Measuring times   & 5 banks, 10 measurements          \\ 
    \bottomrule
    \end{tabular}
\end{table}






\begin{figure}[!t]
	\centering
	\includegraphics[width=3.4in]{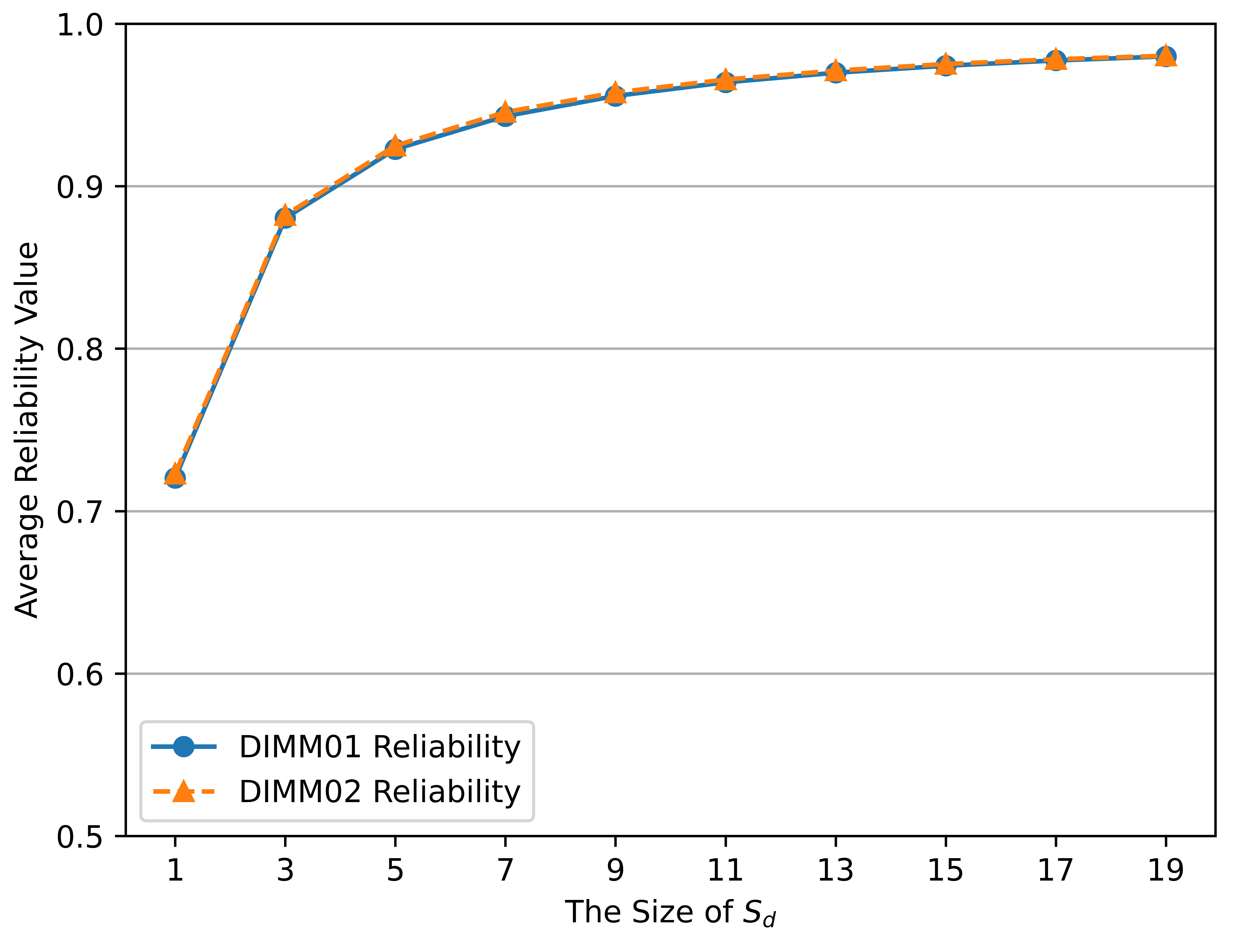}
	\caption{The relationship between the reliability of DRAM fingerprint and the size of $S_d$.}
	\label{S-database-d}
\end{figure}

\textbf{Results.} 
Under the given challenge parameters, we performed 20 fingerprint queries on two DIMMs respectively. 
We arbitrarily select three query results from the 20 fingerprint queries to form the database $S_d$, and then use the remaining single query result as $S_n$.
We tested all possible combinations and got the reliability of the two test DIMMs and the uniqueness between the two DIMMs.
Under the existing test data, the value of uniqueness is 0, the average reliability of $DIMM01$ is 0.88, and the average reliability of $DIMM02$ is also 0.88.
The results show that the distribution difference between uniqueness and reliability is very large, which means that two different devices can be perfectly distinguished and identified.
Fig. \ref{S-database-d} shows how the average value of reliability varies with the number of sets contained in $S_d$.
The result shows that as the physical fingerprint contained in the database $S_d$ of a certain device increases, 
the more physical fingerprint features of the device are collected, the higher the identification accuracy of the device will be.

\begin{figure}[!t]
	\centering
	\includegraphics[width=3.4in]{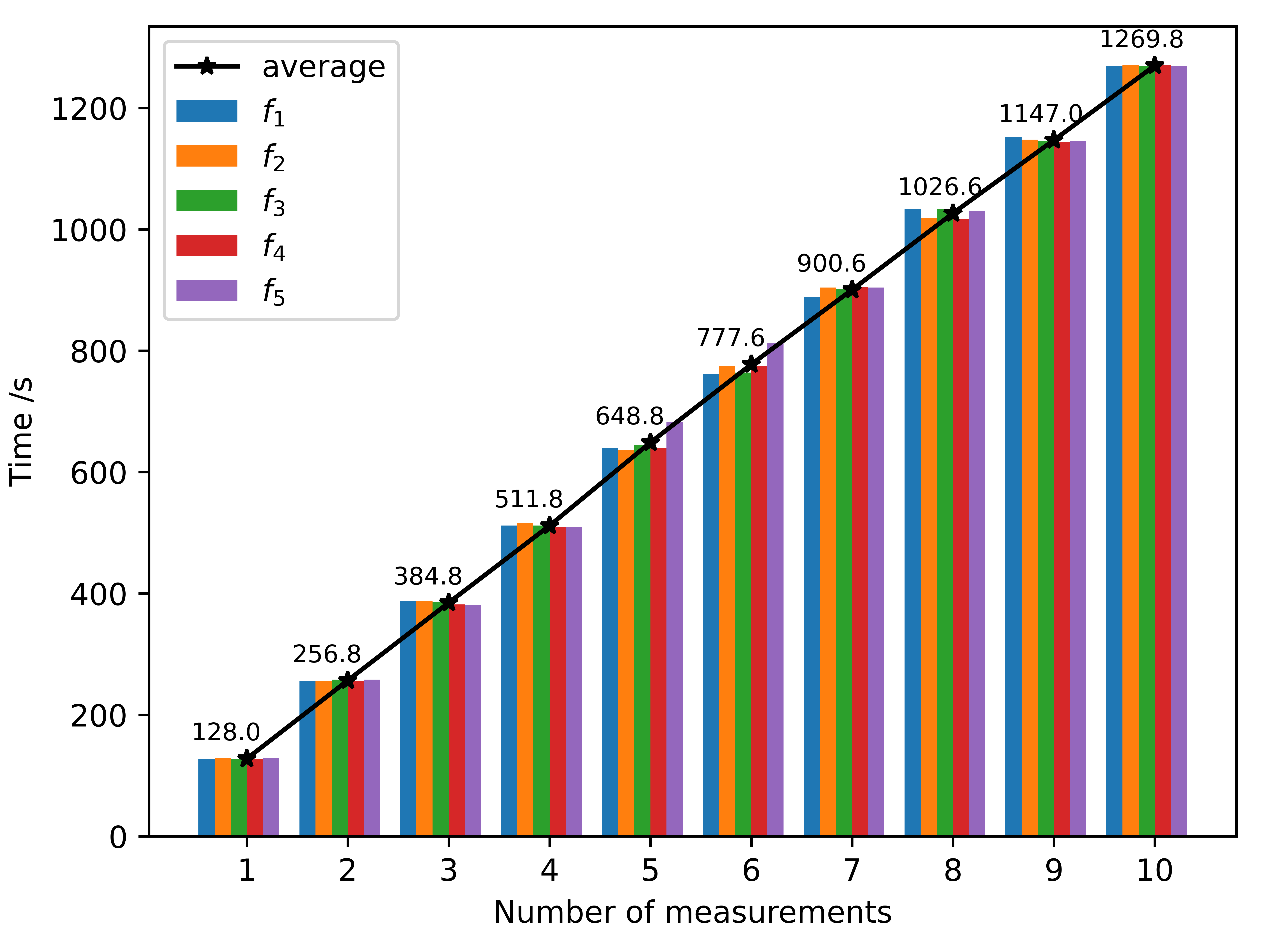}
	\caption{The relationship between the number of measurements in each fingerprint generation process and the fingerprint generation time.}
	\label{mesureing-time}
\end{figure}

\begin{figure}[!t]
	\centering
	\includegraphics[width=3.4in]{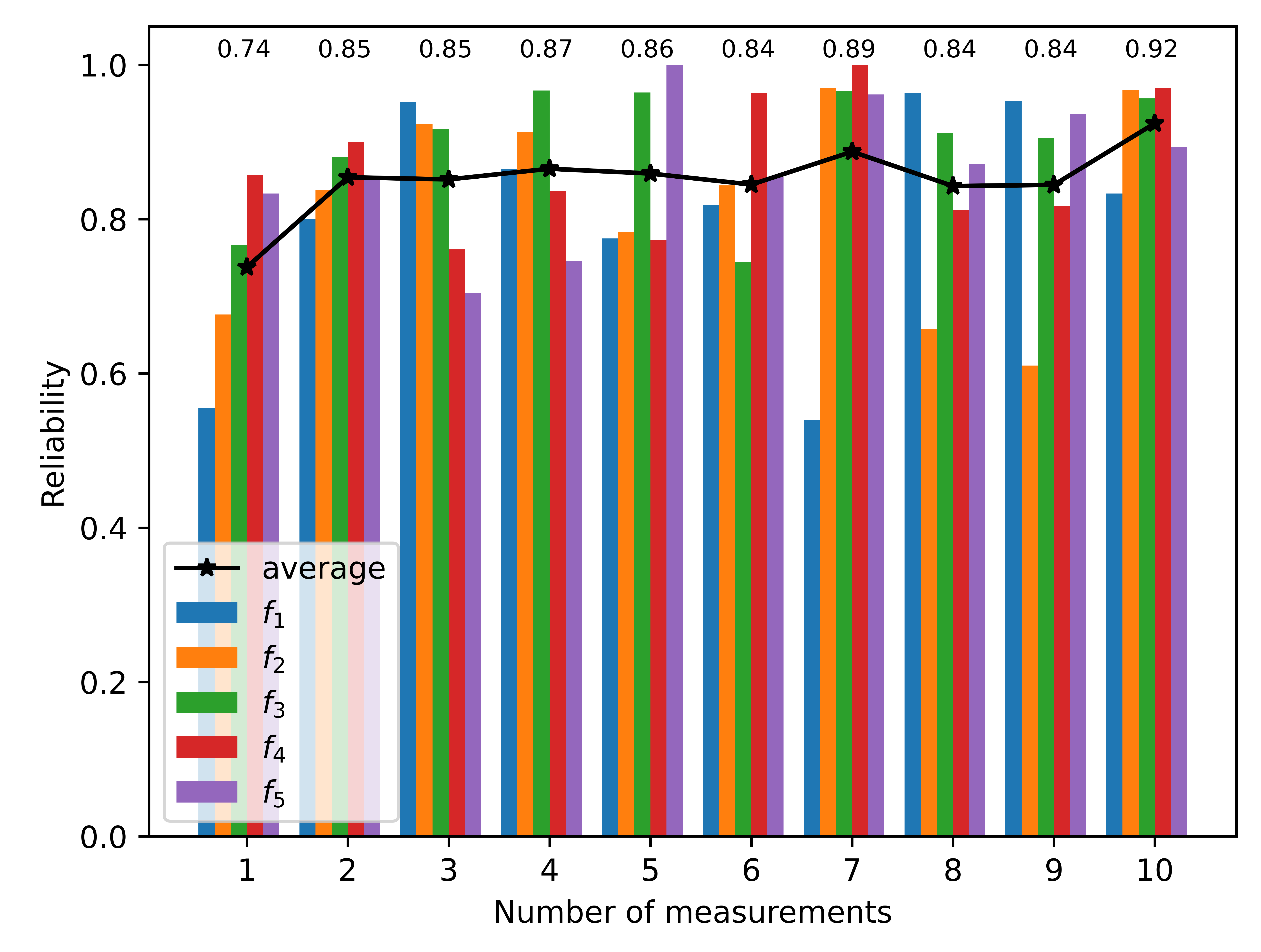}
	\caption{The relationship between the number of measurements and the reliability of the generated fingerprint.}
	\label{mesureing-reliability}
\end{figure}

\textbf{The Time of Fingerprint Generation.} 
The time required for each fingerprint generation process depends on the number of memory rows accessed, the number of banks accessed, and the number of measurements.
The dominant parameter among these parameters is the number of measurements, so we investigate the relationship between the number of measurements and the time of fingerprint generation 
while keeping the other challenge parameters in Table \ref{DRAM-parameters} constant.

As shown in Fig. \ref{mesureing-time}, we conduct five fingerprint queries under different measurement times settings. The bar graph depicts the time required for each fingerprint query, 
and the line graph depicts the average fingerprint generation time for different measurement times settings.
We can see that the fingerprint generation time has a linear relationship with the number of measurements, and the fingerprint generation time under the same number of measurements is relatively stable. 
The average time required to perform one measurement is 128 seconds, and the average time required to perform ten measurements is approximately 1270 seconds.

\textbf{Reliability and the Number of Measurements.}
The more times the measurement is repeated, the longer it takes to generate a fingerprint. But it does not mean that the reliability of fingerprints is higher. 
We want to find the minimum fingerprint generation time required while maintaining reliability that meets device identification needs.
That is, finding the sweet spot between time and reliability.
We use the $Jaccard^{'}$ index to calculate the reliability of the generated fingerprints.
We arbitrarily select one fingerprint set from the five fingerprint query results as $S_n$ in Equation \ref{j_intra}, and the remaining four fingerprint sets form the database $S_d$.

As shown in Fig. \ref{mesureing-reliability}, the reliability values are represented by a bar graph in the figure and a line graph in the figure depicting the average reliability value.
The results show that there is no significant difference in reliability between taking two measurements and taking nine measurements. 
In time-critical scenarios, taking two measurements appears to be a better choice, with a reliability average of 0.85 and an average time required to generate a fingerprint of about 257 seconds.

\subsection{Device Identification Experiment}
The experiments are carried out in a local area network (LAN) environment, and we use one laptop as the detection host and the other eight devices as the target devices to be detected.
The eight target devices used for evaluation are equipped with the same model of CPU and operating system, and some of the devices were equipped with the same DRAM modules.
We detect the target device twice and modify the device's MAC address and IP address before the second detection. 
Therefore, the device can only be identified by the generated physical fingerprint.
The DRAM challenge parameters used to generate the physical fingerprint on the target device are consistent with Table \ref{DRAM-parameters}.

\begin{table*}[!t]
    \caption{The attributes of the target device obtained from the first detection.}
    \label{portraits-1}
    \centering
    \begin{tabular}{l | l l l l}
    \toprule
     & $Device_{A1}$     & $Device_{A2}$       & $Device_{A3}$       & $Device_{A4}$     \\
    \hline
    Physical fingerprint & $<f_{1j}>$, 354 bit flips  & $<f_{2j}>$, 559 bit flips  & $<f_{3j}>$, 1493 bit flips  & $<f_{4j}>$, 51 bit flips  \\
    Memory part number   & M378A1K43DB2-CVF    & M378A1K43DB2-CVF    & M378A1K43DB2-CVF   & M378A1K43DB2-CTD  \\
    Operating system     & \multicolumn{4}{c}{Ubuntu 20.04.3 LTS}        \\
    \toprule
    & $Device_{A5}$     & $Device_{A6}$       & $Device_{A7}$       & $Device_{A8}$     \\
    \hline
    Physical fingerprint & $<f_{5j}>$, 628 bit flips  & $<f_{6j}>$, 2300 bit flips  & $<f_{7j}>$, 272 bit flips  & $<f_{8j}>$, 665 bit flips  \\
    Memory part number   & M378A2K43DB1-CVF    & M378A1K43DB2-CVF    & M378A1K43DB2-CTD   & M378A2K43DB1-CVF  \\
    Operating system     & \multicolumn{4}{c}{Ubuntu 20.04.3 LTS}        \\
    \bottomrule
    \end{tabular}
\end{table*}

\textbf{The First Detection.}
The attributes of target devices obtained from the first detection are shown in Table \ref{portraits-1}.
The attributes of the device consist of physical fingerprint, memory part number, and operating system version.
For each target device, we perform $j$ measurements on the DRAM during the first detection process to collect enough fingerprints of the target device to form a physical fingerprint set $<f_{ij}>$ corresponding to the device $i$.
In addition, we calculate the average number of bit flips over $j$ measurements. The set $<f_{ij}>$ of different devices constitutes the fingerprint dataset $F$, which is used to match the new fingerprints collected subsequently.

\begin{table*}[!t]
    \caption{The attributes of the target device obtained from the second detection.}
    \label{portraits-2}
    \centering
    \begin{tabular}{l | l l l l}
    \toprule
     & $Device_{B1}$     & $Device_{B2}$       & $Device_{B3}$       & $Device_{B4}$      \\
    \hline
    Physical fingerprint & $f_{u1}$, 365 bit flips  & $f_{u2}$, 637 bit flips  & $f_{u3}$, 432 bit flips  & $f_{u4}$, 2829 bit flips  \\
    Memory part number   & M378A1K43DB2-CVF    & M378A2K43DB1-CVF    & M378A1K43DB2-CVF   & M378A1K43DB2-CVF  \\
    Operating system     & Ubuntu 18.04.3 LTS & \multicolumn{3}{c}{Ubuntu 20.04.3 LTS}        \\
    \toprule
    & $Device_{B5}$     & $Device_{B6}$       & $Device_{B7}$       & $Device_{B8}$     \\
    \hline
    Physical fingerprint &  $f_{u5}$, 1600 bit flips  &  $f_{u6}$, 456 bit flips  &  $f_{u7}$, 295 bit flips  &  $f_{u8}$, 40 bit flips \\
    Memory part number   & M378A1K43DB2-CVF    & M378A2K43DB1-CVF    & M378A1K43DB2-CTD   & M378A1K43DB2-CTD  \\
    Operating system     & \multicolumn{4}{c}{Ubuntu 20.04.3 LTS}        \\
    \bottomrule
    \end{tabular}
\end{table*}

\textbf{The Second Detection.} 
We modify the MAC address and IP address of the eight target devices and reinstall the operating system of one of the devices, and then perform the second detection.
The attributes of target devices obtained from the second detection are shown in Table \ref{portraits-2}.
We identify these eight devices as $B_1$, $B_2$, ..., and $B_8$, respectively. For each device, we only measure the DRAM once and collect the device's corresponding physical fingerprint $f_u$.
Our goal is to identify the correspondence between the newly detected target device and the previously detected device. 
In other words, even if the target device has replaced its virtual attributes such as MAC address and IP address, we can still identify the target device through its unique physical fingerprint.

\textbf{Results.}
We match the devices in the two detection results according to the proposed device identification Algorithm \ref{device-identification-alg}. 
That is, the newly collected fingerprint $f_u$ is matched with the device fingerprint in the dataset $F$. 
We instantiate the \textit{FingerprintMatch} function in Algorithm \ref{device-identification-alg} with $Jaccard~index$ for the initial matching of fingerprint $f_u$ and device fingerprint $f_{i1}$. 
Then we instantiate the \textit{GetSimilarity} function in the Algorithm \ref{device-identification-alg} with $Jaccard^{'}~index$, which is used to calculate the similarity of the fingerprint $f_u$ to the candidate device fingerprint set $<f_{ij}>$.

\begin{figure}[!t]
	\centering
	\includegraphics[width=0.9\linewidth]{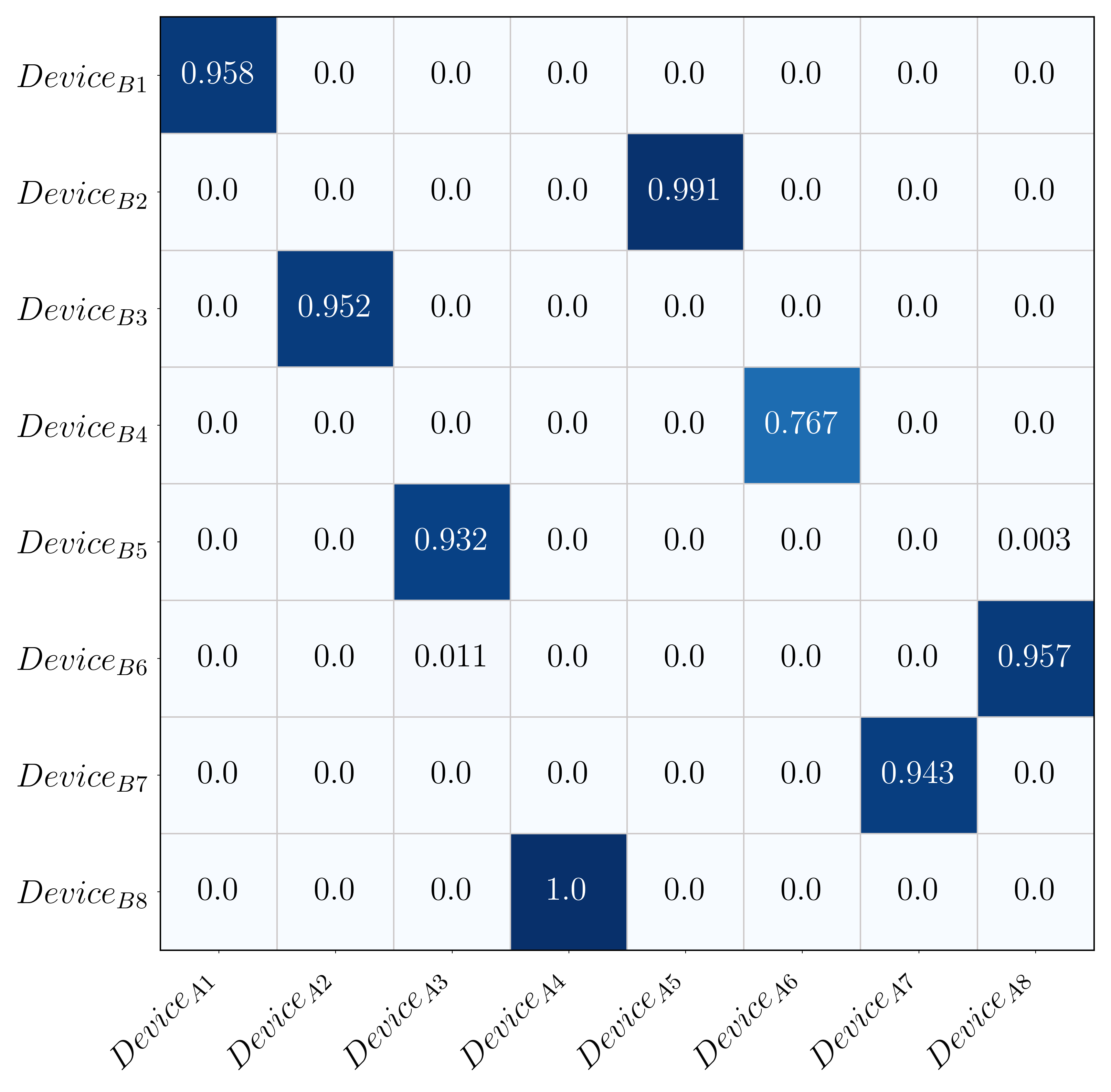}
        \caption{$Jaccard^{'}$ results for device identification.}
	\label{device-match}
\end{figure}

The device fingerprint similarity results calculated according to the $Jaccard^{'}~index$ are shown in Fig. \ref{device-match}.
The results show that although device $A_1$ has reinstalled the operating system and changed its IP address and MAC address, we can still identify device $A_1$ by physical fingerprint.
Specifically, the device $B_1$ is the previous device $A_1$. In addition, device $B_5$ has a high probability of being the previous device $A_3$, not device $A_8$; device $B_6$ has a high probability of being the previous device $A_8$, not device $A_3$.
The similarity value of devices $B_8$ and $A_4$ in Fig. \ref{device-match} is 1, which is reasonable. The reason is that the bit flip locations contained in the fingerprint $f_{u8}$ of the device $B_8$ are also all in the fingerprint set $<f_{4j}>$ of the device $A_4$.

\begin{figure}[!t]
	\centering
	\includegraphics[width=0.9\linewidth]{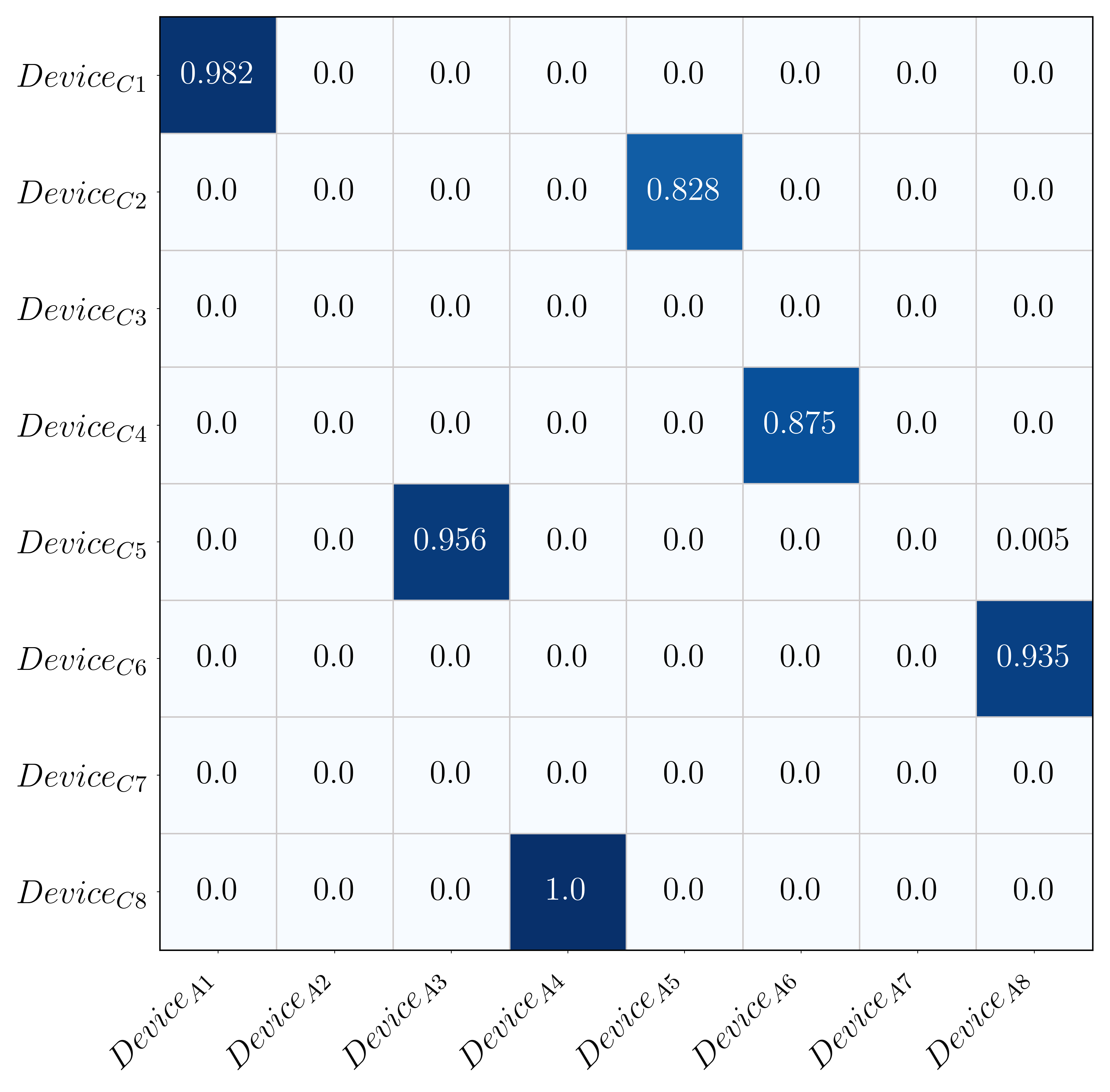}
        \caption{$Jaccard^{'}$ results for new device identification.}
	\label{device-match-1}
\end{figure}

\textbf{New Device Identification.}
Based on the above experiments, we replace two of the eight devices with new ones and left the other six unchanged.
All virtual attributes of the two new devices are the same as those in the second detection experiment they replaced, consistent with Table \ref{portraits-2}.
Our goal is to identify two new devices using the device's physical fingerprint.
We perform a new detection of the target device and identify the eight devices as $C_1$, $C_2$, ..., and $C_8$, respectively.
The fingerprint similarity results calculated according to the $Jaccard^{'}~index$ are shown in Fig. \ref{device-match-1}.
According to the results in Fig. \ref{device-match-1}, we can know that $C_3$ and $C_7$ are newly added devices. 
The identification relationships of other devices are the same as before.

\subsection{One DIMM on Multiple Computers}
\textbf{Motivation.}
We want to figure out whether the generated physical fingerprint is uniquely bound to the DRAM or is bound to the entirety of the DRAM and the computer it resides in.
If it is the former, it means that when a certain DIMM runs on two different devices in different time periods and is detected respectively in these two time periods, the two devices will be identified as the same device. 
Because the physical fingerprints of the two devices are generated based on the same DIMM.
If it is the latter, in this case, the two devices will be identified as two different devices. In this way, the identification of the device will be more accurate.

To investigate this question, we conduct the following experiment.
We use the same DIMM to run on three different computers $A$, $B$, and $C$, and get three sets of physical fingerprints. 
The configurations of $A$ and $B$ are exactly the same, and $C$ is a computer with a different brand and configuration from $A$ and $B$.
The device fingerprint similarity results calculated according to the $Jaccard^{'}$ index are shown in Table \ref{one-dimm}.

\begin{table}[!t]
    \caption{The similarity of fingerprints obtained by different computers based on the same DIMM.}
    \label{one-dimm}
    \centering
    \begin{tabular}{c | c c c || c | c c c}
    \toprule
    \makecell{$DIMM01$} & $A$ & $B$ & $C$ & \makecell{$DIMM02$} & $A$ & $B$ & $C$ \\
    \hline
    $A$ & 1 & 0 & 0 & $A$ & 1 & 0 & 0 \\
    $B$ & 0 & 1 & 0 & $B$ & 0 & 1 & 0 \\
    $C$ & 0 & 0 & 1 & $C$ & 0 & 0 & 1 \\
    \bottomrule
    \end{tabular}
\end{table}

\begin{table}[!t]
    \caption{Average number of bit flips per fingerprint generation.}
    \label{average-bit-flips}
    \centering
    \begin{tabular}{c | c c c}
    \toprule
    $Bit~Flips$ & $Device_A$ & $Device_B$ & $Device_C$ \\
    \hline
    $DIMM01$ & 352 & 281 & 242 \\
    $DIMM02$ & 559 & 405 & 124 \\
    \bottomrule
    \end{tabular}
\end{table}

\textbf{Results.}
The results show that even if the same DIMM runs on different devices, the device fingerprints generated by different devices based on the same DIMM are different, and these devices will not be identified as the same device.
Therefore, we conclude that the generated physical fingerprint is bound to the whole consisting of the DRAM and the computer in which it resides rather than only to the DRAM module.
We also count the average number of bit flips during fingerprint generation for the same DIMM on different devices $A$, $B$, and $C$.
The results in Table \ref{average-bit-flips} show that the same DIMM gets different numbers of bit flips on different devices.
This means that there is a difference in the average number of bit flips for the same DIMM, whether it is between devices of the same brand and configuration or between devices of different brands and configurations.

We think a reasonable explanation is that the overall environment of the host computer where the DIMM is located is also an important parameter that constitutes the DRAM challenge.
Although the other challenge parameters are the same, the DRAM challenge is not the same due to differences in operating voltage, BIOS settings, CPU cache mode, address mapping relationship, etc.
As a result, the physical fingerprints obtained are different.

\section{Related Work}

\subsection{Device Fingerprinting}
The device fingerprinting technique forms a device fingerprint by extracting the hardware or software characteristics of the device to identify, track, and authenticate the device or user 
\cite{alaca2016device, zhang2019sensorid, laor2022drawnapart, sanchez2018clock, kohno2005remote, das2014you, cao2017cross}.
At present, there are two mainstream methods: one is to form fingerprints based on the attributes of the browser \cite{eckersley2010unique, vastel2018fp}, and the other is to form fingerprints based on the hardware characteristics of the device. 
The advantage of hardware fingerprint over browser fingerprint is that it is generally more stable. 
Hardware will inevitably introduce some random noise during the manufacturing process, resulting in differences in certain characteristics of hardware with the same design and the same specifications. 
Therefore, the physical fingerprint of the device can be formed based on some hardware modules of the device.


One technique is to identify devices based on the timing differences in the device's execution of a sequence of instructions \cite{sanchez2018clock, kohno2005remote}. 
The reason for this difference in execution time is that crystal-based oscillators cause differences in the clock frequency of the device.
For example, extract hardware fingerprints based on system clock skew and CPU characteristics.
Kohno et al. \cite{kohno2005remote} proposed a remote clock skew estimation technology that uses TCP and ICPM timestamps to identify physical devices.
Sanchez-Rola et al. \cite{sanchez2018clock} proposed a method to calculate hardware fingerprints by measuring the execution time of the CPU to run a specific instruction sequence.

There is also work to extract fingerprints of hardware through GPU.
Li et al. \cite{li2015poster} exploit the inherent randomness of the GPU to generate a unique, GPU-specific signature.
Laor et al. \cite{laor2022drawnapart} construct device fingerprints based on the time difference between execution units (EUs) of a GPU when drawing operations are performed.
In contrast to these works above, the physical fingerprints generated by our proposed DRAM-based device identification technique are not affected by software layers such as the operating system. 
Moreover, the generated fingerprint is relatively more stable, which improves the accuracy of device identification.

\subsection{DRAM PUFs.}
Physically Unclonable Functions (PUFs) provide unique fingerprints to physical entities through randomness introduced during the manufacturing process.
Traditional PUFs based on DRAM characteristics include decay-based DRAM PUFs \cite{tehranipoor2016dram, schaller2018decay} and latency-based DRAM PUFs \cite{kim2018dram, miskelly2020fast}.
Schaller et al. \cite{schaller2017intrinsic, anagnostopoulos2018intrinsic} presented the work to design PUF leveraging the Rowhammer effect. 
They implemented Rowhammer PUF on a PandaBoard equipped with DDR2 memory and tested the performance of PUF under different conditions.
However, they have only been evaluated on an experiment board such as the PandaBoard, which is not practical. 
In contrast, instead of using Rowhammer to form the PUF, our evaluation scenario is fingerprinting of PC devices equipped with DDR4 memory, which is more difficult and more practical.

PUF requires the generation of a large number of stable challenge-response pairs (CRPs) and is often used in authentication protocols.
However, Zeitouni et~al. \cite{zeitouni2018s} achieved a modeling attack on Rowhammer PUF by collecting a large number of CRPs exchanged in the authentication protocol, which makes Rowhammer PUF unsuitable as a PUF to be used in the authentication protocol. 
Because the authentication protocol needs to exchange CRPs to achieve authentication, the attacker can collect the CRPs to model the PUF.
In contrast, our Rowhammer-based fingerprinting technique does not require CRP exchanges, the PUF attacker cannot obtain our DRAM challenge parameters and generated fingerprints, so this attack is ineffective.

\section{Conclusion}
In this paper, we introduce FPHammer, a new technique for device identification based on DRAM fingerprinting. 
We present the detailed design flow for extracting physical fingerprints on personal computers equipped with DDR4 memory.
Furthermore, we design a device identification algorithm to link fingerprints belonging to the same device to the same device $id$.
The evaluation results show that the generated fingerprints have good properties and can be used to distinguish two DIMMs with the same configuration.
We also conduct a series of experiments to evaluate the performance of FPHammer.
The experimental results show that even if the device replaces virtual attributes such as MAC address and IP address or even reinstalls the operating system, FPHammer can still identify the target device.
Furthermore, the generated physical fingerprint is bound to the entire device rather than only the DRAM module.
This ensures unique identification of the device, even if different devices use the same DRAM module at different times.

\section*{Acknowledgment}
This work is supported by the National Key R\&D Program of China (2021YFB2700200), 
the Natural Science Foundation of China 62372022, 62002006, U2241213, U21B2021, 62172025, 61932011, 61932014, 61972018, 61972019, 61772538, 32071775, 91646203, 
the Defense Industrial Technology Development Program JCKY2021211B017.


\bibliographystyle{IEEEtranS}
\bibliography{references}


\end{document}